\documentclass[a4paper,superscriptaddress,english,twocolumn,showkeys]{revtex4-1} 
\usepackage{times}
\usepackage{graphicx}
\usepackage{amssymb}
\usepackage{hyperref} 

\begin{document}
	
\title{\textbf{New fabrication technique for highly sensitive qPlus sensor with well-defined spring constant}}

\author{Hatem Labidi}
\email{hatem.labidi@ualberta.ca}
\affiliation{Department of Physics, University of Alberta, Edmonton, Alberta,T6G 2J1, Canada}
\affiliation{National Institute for Nanotechnology, National Research
	Council of Canada,Edmonton, Alberta, T6G 2M9, Canada}

\author{Martin Kupsta}
\affiliation{National Institute for Nanotechnology, National Research
	Council of Canada,Edmonton, Alberta, T6G 2M9, Canada}

\author{Taleana Huff}
\affiliation{Department of Physics, University of Alberta, Edmonton, Alberta,T6G 2J1, Canada}
\affiliation{National Institute for Nanotechnology, National Research
	Council of Canada,Edmonton, Alberta, T6G 2M9, Canada}

\author{Mark Salomons}
\affiliation{National Institute for Nanotechnology, National Research
	Council of Canada,Edmonton, Alberta, T6G 2M9, Canada}

\author{Douglas Vick}
\affiliation{National Institute for Nanotechnology, National Research Council of Canada,Edmonton, Alberta, T6G 2M9, Canada}

\author{Marco Taucer}
\affiliation{Department of Physics, University of Alberta, Edmonton, Alberta,T6G 2J1, Canada}
\affiliation{National Institute for Nanotechnology, National Research
	Council of Canada,Edmonton, Alberta, T6G 2M9, Canada}

\author{Jason Pitters}
\affiliation{National Institute for Nanotechnology, National Research
	Council of Canada,Edmonton, Alberta, T6G 2M9, Canada}

\author{Robert A. Wolkow}
\affiliation{Department of Physics, University of Alberta, Edmonton, Alberta,T6G 2J1, Canada}
\affiliation{National Institute for Nanotechnology, National Research
	Council of Canada,Edmonton, Alberta, T6G 2M9, Canada}

\begin{abstract}
	A new technique for the fabrication of highly sensitive qPlus sensor for atomic force microscopy (AFM) is described. Focused ion beam was used to cut then weld onto a bare quartz tuning fork a sharp micro-tip from an electrochemically etched tungsten wire. The resulting qPlus sensor exhibits high resonance frequency and quality factor allowing increased force gradient sensitivity. Its spring constant can be determined precisely which allows accurate quantitative AFM measurements. The sensor is shown to be very stable and could undergo usual UHV tip cleaning including e-beam and field evaporation as well as in-situ STM tip treatment. Preliminary results with STM and AFM atomic resolution imaging at $4.5\,K$ of the silicon $Si(111)-7\times 7$ surface are presented. 
\end{abstract}


\maketitle
\date{\today}

\section{Introduction}
Since the sixties, quartz tuning forks (QTF) have been widely used in watches and other electronic devices as frequency standard mainly because of the stability of their mechanical properties with temperature fluctuations\cite{Giessibl.1998,Giessibl.2009}. More recently, their high resonance frequency, quality factor and spring constant as well as their self sensing capacity, thanks to the piezoelectricity of quartz, motivated their implementation in many scanning probe microscopy techniques such as scanning near field microscopy\cite{Karrai.1995,Ping.1998}, scanning near-field acoustic microscopy\cite{Gunther.1989} and magnetic force microscopy\cite{Edwards.1997,Todorovic.1998}. This has made QTF based sensors a very important tool for different scientific communities\cite{Edwards.1997}.

QTF were also implemented in atomic force microscopy (AFM) to serve as both actuator and sensor for tip-sample interactions which eliminated the need for optics and allowed low oscillation amplitude operations\cite{Edwards.1997}. However, the gluing of a metallic tip to the QTF and the interactions with the surface induced break of the QTF symmetry resulting in low scanning speed and resolution. The qPlus sensor (QPS) introduced by Giessibl\cite{Giessibl.2000} gave a simple and efficient solution for these problems. In this design, one of the QTF prongs is firmly fixed to a supporting structure (ceramic) and a metallic tip (usually tungsten) is glued to the free prong using epoxy\cite{Giessibl.1998,Giessibl.2000}. In this configuration, only one prong is deflected during scanning of a surface and the QTF is behaving essentially as a self sensing cantilever which makes fast scanning possible and easy to interpret\cite{Giessibl.1998,Giessibl.2000}. Since its introduction, the QPS attracted increasing attention and atomic resolution at low temperature in non contact AFM (NC-AFM) mode is now routinely achieved by different groups\cite{Giessibl.2000,Gross.2009,Sharp.2012}.

Despite their wide use and commercialization, QPS are still handmade and suffer from decreased resonance frequency and quality factor when compared to the bare QPS (without tip) due to the mass load induced by the relatively large and heavy tungsten tip fixed to the free prong using conductive epoxy\cite{Giessibl.2000,Giessibl.2009,Berger.2013,Falter.2014}. Moreover, it's impossible to know exactly the added mass, the amount of epoxy used and the position of the tip on the free prong. These imprecision result in a large spread of the mechanical properties from one fabricated sensor to another and makes it very difficult to calibrate the spring constant (stiffness)\cite{Falter.2014}. Even the type of epoxy used was shown to induce significant difference\cite{Vorden.2012}.

In addition to demonstrating sub-molecular resolution imaging and atomic/molecular manipulation, recent developments in NC-AFM techniques using QPS introduced quantitative interpretation of atomic forces\cite{Ternes.2008,Albers.2009,Gross.2009,Gross.2009-2,Langewisch.2013,Weymouth.2014,Leoni.2011}. However, the exact value of the sensor's stiffness ($k$) is of critical importance to convert the experimentally measured frequency shift ($\Delta f$) to interaction forces ($F$) and also to calculate other important physical quantities such as energy dissipation\cite{Giessibl.2001,Sader.2004,Falter.2014}. Since $k$ can't be known exactly for standard QPS, these quantitative measurements lack  precision, have large errors and would be different from one sensor to another\cite{Giessibl.2009,Falter.2013,Falter.2014}. This issue motivated recent efforts by different groups in finding techniques for the calibration of the QPS spring constant\cite{Vorden.2012,Berger.2013,Falter.2014}. However, these studies concluded to the difficulty of this task due to the imprecise fabrication method itself e.g. large tip mass, variability in epoxy amount/type, and tip positioning on the QTF.

\begin{figure*}[ht]
	\centering
	\includegraphics[width=\textwidth]{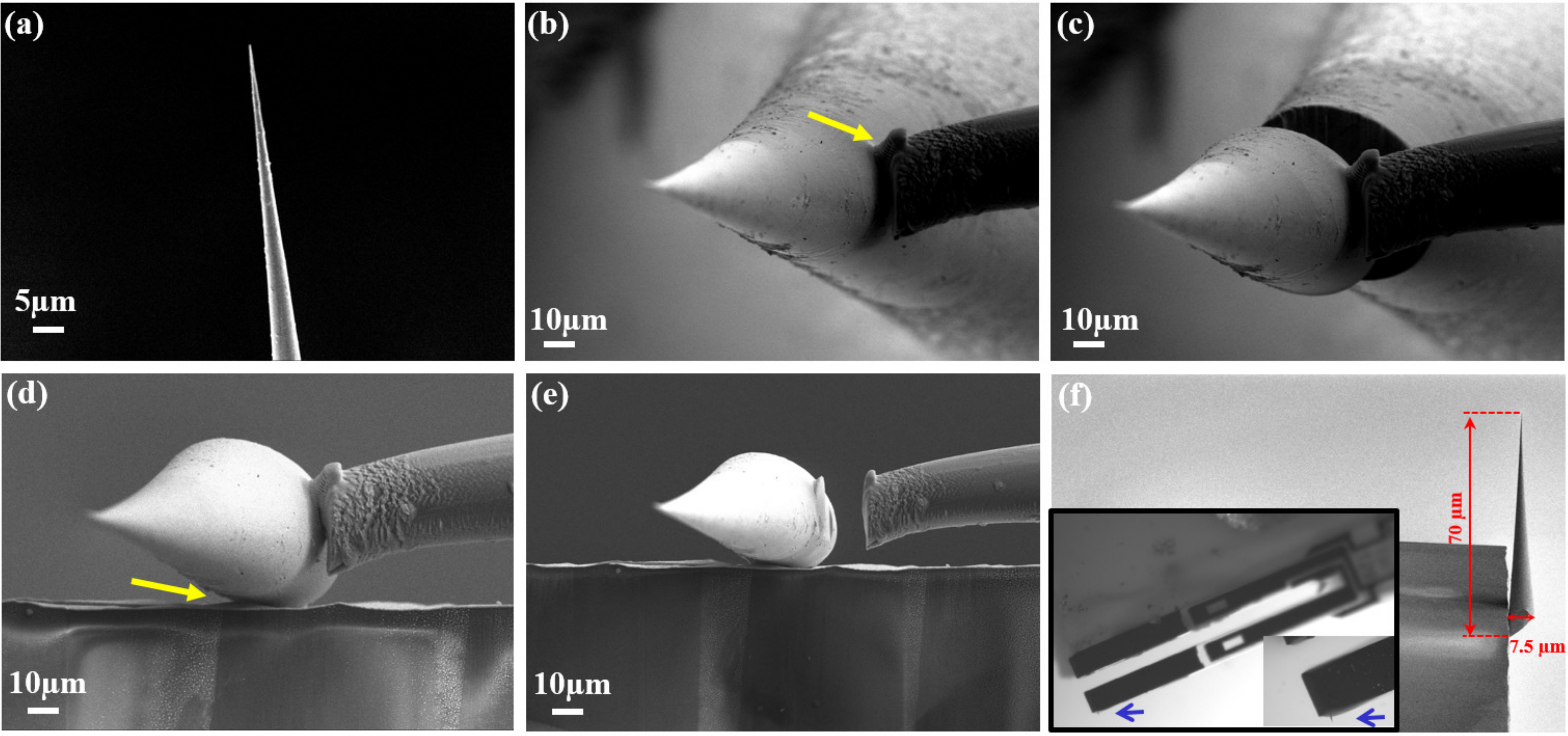}
	\caption{Series of SEM images depicting the fabrication process of mounting a tungsten micro-tip onto a QPS (see text for details): (a) Tungsten tip selection. (b) Welding the micro manipulator to the tungsten tip. (c) Micro-tip detachment after FIB cutting. (d) Placing then welding of the FIB cut micro-tip on the QPS. (e) Detachment of the micro manipulator and leaving the micro-tip fixed to the QPS. (f) SEM image showing clearly dimensions and shape of the micro-tip. The insert in (f) is an optical micrograph of the hole QPS after FIB fabrication.}
	\label{Fig1}
\end{figure*}

In this letter, we describe a new technique that uses focused ion beam (FIB) for the precise and controllable fabrication of highly sensitive and robust epoxy free QPS. The sensor shows almost no change in the resonance frequency and quality factor if compared to the bare QPS. Moreover, we demonstrate that the spring constant of such sensors can be determined precisely which is critical for accurate quantitative AFM measurements.

\section{Fabrication method}
Several studies in the literature already mentioned the use of FIB to optimize QTF based force sensors\cite{Hida.2008,Gross.2012,Sharp.2012}. In our approach, we make use of the unique capabilities of FIB as a micro-engineering tool in all the fabrication steps. A Hitachi NB5000, a dual beam FIB/SEM, microscope was used. The FIB uses Gallium as its liquid metal ion source and is capable of accelerating voltages between $1\,kV$ (polishing and imaging) to $40\, kV$ (milling). Two holders were used in the fabrication process, one to carry tips and the other to carry QPSs. Since the FIB column on the NB5000 is vertically mounted and the SEM mounted $58^{\circ}$ from the vertical, the stage needs to be tilted in order to cut the tungsten tip from its base and then to mount the cut part of the tip perpendicularly onto the QPS. A micro manipulator is also present on the microscope. An in-situ tungsten ($W(CO)_6$) deposition process\cite{Utke.2008} ensuring mechanical stability and electrical conductivity is used to weld the micro-tip to the conductive side of the QPS. To cut through the bulk for the tip, ion beam currents of $64\, nA$ were used. For finer milling and welding, beam currents of $1-0.1 \, nA$ were used. Images were captured using the SEM at $5\, kV$.

Tungsten tips suitable for STM were electrochemically etched from a $0.25\,mm$ diameter polycrystalline wire. The etching parameters were optimized to obtain a sharp tip with a radius of curvature of less than $25\,nm$ as seen in figure \ref{Fig1}-a. Tungsten was chosen mainly because of the possibility to easily and routinely obtain sharp tips with smooth and well defined conic geometry\cite{Rezeq.2006}. However, as far as FIB is concerned, any other material can be used as tip\cite{Utke.2008}. Usually, several tips are etched then mounted on a same holder and loaded in the FIB/SEM microscope. 

After examination with SEM, a sharp tip is selected (figure \ref{Fig1}-a) then tilted to $58^{\circ}$. The micro manipulator is introduced into the sample chamber and welded onto the side of the tip at the desired length (figure \ref{Fig1}-b). 
FIB is then used to cut the tip at the edge of the welded area leaving a micro-tip held only by the probe (figure \ref{Fig1}-c). The micro manipulator and tips are then retracted from the system and the QPS introduced. During the fabrication processes, care is taken not to image the tip apex with the Gallium beam.

Since the plucked tip is tilted at $58^{\circ}$, the QPS needs to be tilted as well to allow for perpendicular mounting of the tip to the patterned gold side of the QPS free prong (figure \ref{Fig1}-f). Thanks to the use of the micro manipulator, it is possible to place the tip with sub-micron accuracy. This is an important feature as it allows choosing the effective length of the prong and hence the stiffness of the sensor\cite{Falter.2014}. After firmly welding the micro-tip to the conductive gold patterned side of the QPS free prong, it is freed from the micro-manipulator (figure \ref{Fig1}-g). Using FIB beam, it is possible to clean the tip from welding and probe residues. We can note here that the whole FIB fabrication process can be relatively fast: about 1 hour if optimized.

As seen in figure \ref{Fig1}-g, the micro-tip dimensions can be determined directly from SEM images after tilt correction. In order to minimize errors, we used a computer-aided design (CAD) program (Inventor) to calculate the volume of the micro tip. The mass can then be estimated considering tungsten density $\rho_{w}=19.3\,g \, cm^{-3}$). This method yields $m=(17.7\pm0.8)\, ng$ in the example of figure \ref{Fig1}-g, where an error of 5\% on the volume determination was considered. Since the amount of deposited tungsten used to weld the micro-tip is negligible\cite{Utke.2008}, the estimated micro-tip mass can be considered as the total mass load to the bare QPS. 

\section{Characterization of the mechanical properties of the new sensor}
\subsection{High resonance frequency and quality factor}
To rapidly characterize the mechanical properties of the QPS, we built a simple test unit compatible with the commercial omicron sensors\cite{Bettac.2009,Berger.2013}. As shown in figure \ref{Fig2}-a, the QPS in this setup is mechanically excited by a piezoelectric motor. We used a signal analyzer (Stanford Research Systems SR780) to deliver the excitation signal. The qPlus signal is then amplified (Femto DLPCA-200 preamp) prior to feeding it to the signal analyzer. The oscillation amplitude of the qPlus can then be recorded as a function of the swept excitation frequency. 

Figure \ref{Fig2}-b shows the resonance curve for the same QPS before (black curve) and after (red curve) FIB welding of the micro-tip. As expected, the minimal mass load induced by our new fabrication technique ensures a high Q factor and very little frequency shift (less than $5\, Hz$ in this example) when compared to the bare sensor (without tip) which is very important for increased force detection. This compares with the commercially available QPS, on an identical cantilever, where the resonance frequency drops dramatically after tip gluing with typical values ranging from $23\, kHz$ to $26\, kHz$.  Also, as demonstrated in a recent study by Tung et al\cite{Tung.2010}, minimal mass load is of critical importance for stable oscillation of qPlus in higher-order eigenmodes. 

\subsection{A well defined spring constant}
\begin{figure}[ht]
	\centering
	\includegraphics[width=\columnwidth]{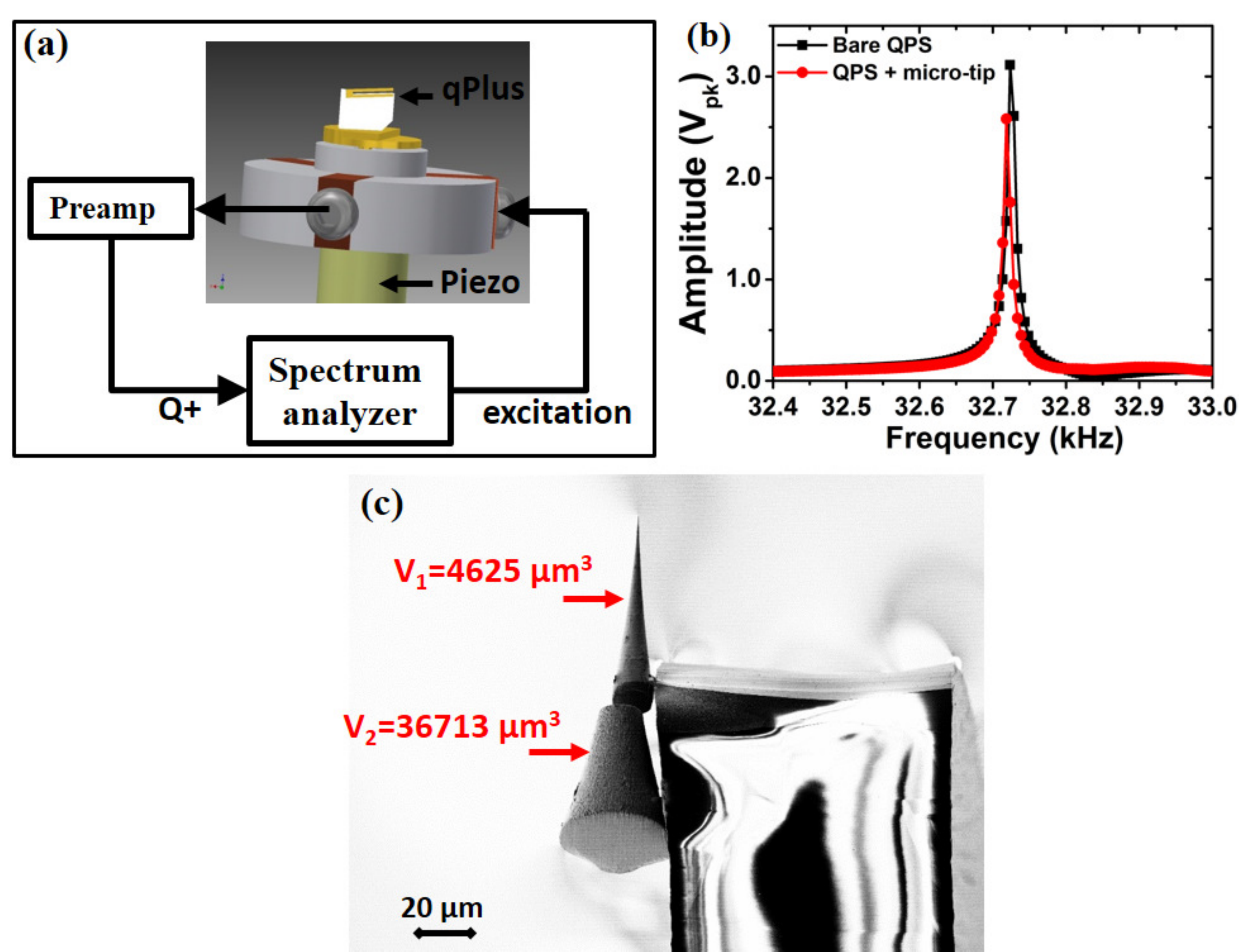}
	\caption{(a) Diagram of the experimental setup used to characterize the mechanical properties of QPS in different vaccuum conditions. (b) Resonance curve of the same QPS before (blue curve) and after (red curve) the FIB welding of the tungsten micro tip. (c) SEM image of the QPS with a micro-tip and an additional mass showing the calculated volumes.}
	\label{Fig2}
\end{figure}

For silicon cantilevers, it was established that the spring constant ($k$) can be accurately determined experimentally by measuring the resonance frequency of the sensor before ($\nu_0$) and after ($\nu_1$) the addition of a well known mass $M$, following equation \ref{eq1}:
\begin{equation} \label{eq1}
k=(2\pi)^2\frac{M}{\left( \frac{1}{\nu_1}\right)^2-\left(\frac{1}{\nu_0}\right)^2}
\end{equation}
This added mass method, usually referred to as the Cleavland method\cite{Cleveland.1993,Gibson.2001}, was recently used to calibrate the stiffness of QPS\cite{Berger.2013,Falter.2014,Kim.2014}.
Using our setup (\ref{Fig2}-a), the resonance frequency can be easily measured with good precision (less than 0.05\% error) whereas the added mass can be estimated from the SEM images\cite{Cleveland.1993,Berger.2013,Kim.2014} as explained in section 2.

Figure \ref{Fig2}-c shows a QPS with a micro-tip and an additional larger tungsten mass (cut from the same tip). We measured resonance frequencies of $32552\, Hz$, $32522\, Hz$ and $32285\, Hz$ for the bare QPS, the QPS with micro-tip and the QPS with additional mass load respectively.
Equation \ref{eq1} yields $k_0=(2025\pm 45)\, Nm^{-1}$ for the bare QPS and $k_1=(2005\pm 50)\, Nm^{-1}$ the QPS with micro-tip. This clearly shows that unlike the standard qPlus sensor, the FIB controlled welding of the micro-tip induces almost no change to the spring constant of the bare QPS.

To theoretically estimate the spring constant, the beam formula as in equation \ref{eq2} can be used\cite{Berger.2013,Falter.2014}:
\begin{equation}\label{eq2}
k=\frac{E\, w}{4} \left( \frac{t}{\Delta L} \right) ^3
\end{equation}
where E is Young’s elastic modulus ($ E = 78.7\, GPa$ for quartz), t
is the thickness ($0.216\, mm$), w is the width ($0.120\, mm$) and $\Delta L$ ($2.282\, mm$) is the effective length of the prong (beam)\cite{Falter.2014}. 
This gives $k_{th}=2002\, Nm^{-1}$ for the QPS used in the example of figure \ref{Fig2}-c, in good agreement with the experimental result. Since no epoxy is used in our fabrication process and the amount of tungsten weld is negligible, the stiffness won't change with temperature as in the case of the regular hand made sensors\cite{Vorden.2012}. Therefore, the spring constant determined theoretically using equation \ref{eq2} will be almost the same during NC-AFM experiments at cryogenic temperatures. We can take into account the small rise of 1\% in Young's modulus of quartz at $5\,K$\cite{Hembacher.2002}. It must be noted here that according to a recent study\cite{Kim.2014}, the beam formula would not be a good approximation to the experimental stiffness for QPS with larger dimensions.

\section{Low temperature STM and NC-AFM experiments}
\begin{figure}[ht]
	\centering
	\includegraphics[width=0.95\columnwidth]{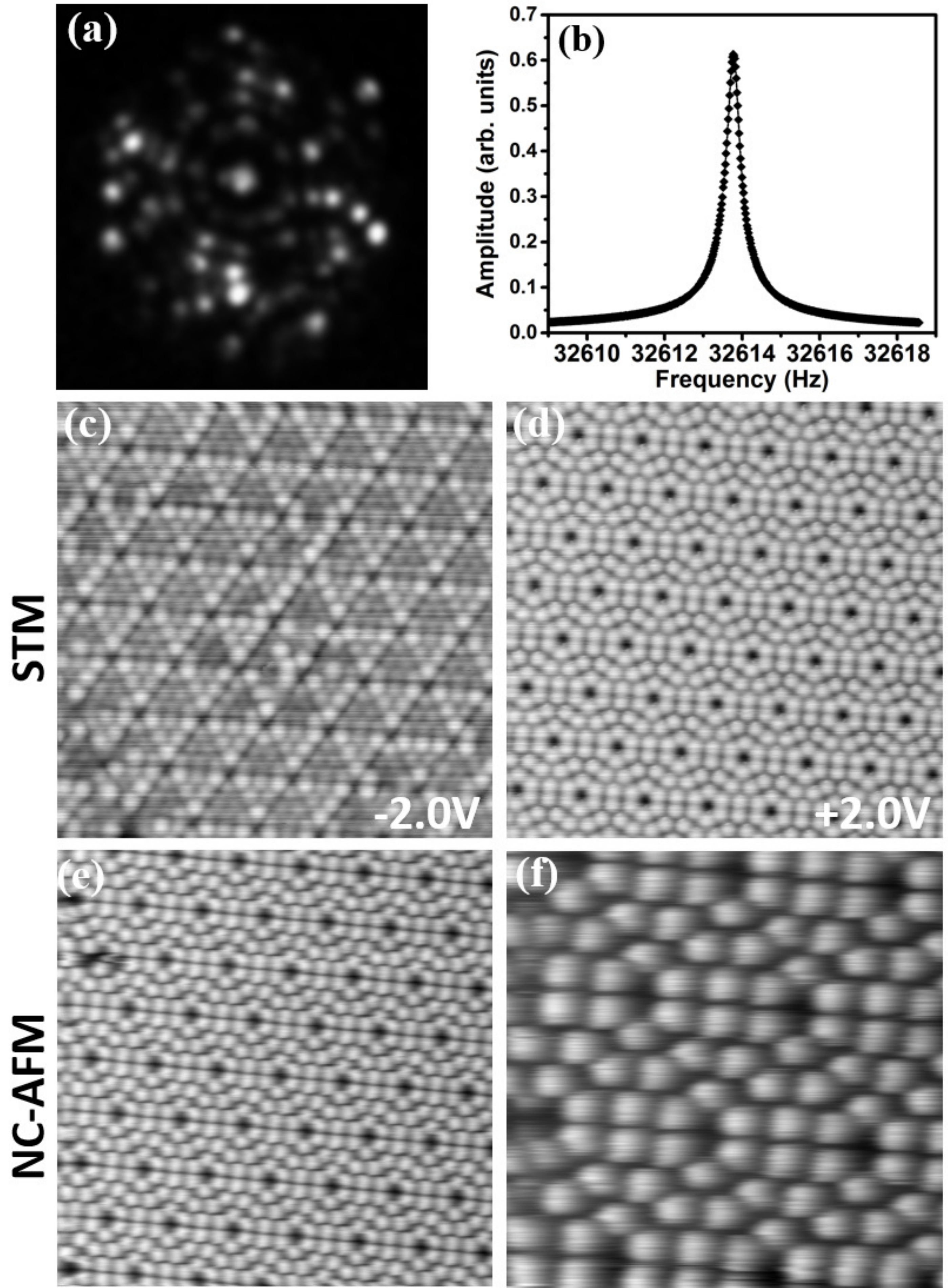}
	\caption{(a) FIM image of the tungsten micro tip mounted on the QPS. (b) Resonance curve of the QPS at 4.5K: $f_0=32613.8\, Hz$ and $Q=95500$. (c) and (d): ($(20\times 20)\, nm^2$) constant current ($30\, pA$) images of the  STM images of the $Si(111)-7\times 7$ surface in filled and empty states respectively.  $(20\times 20)\, nm^2$ (e) and $(8\times 8)\, nm^2$ (f) NC-AFM constant frequency ($-12\, Hz$) images at $0\, V$ with oscillation amplitude of $500\, pm$ and scan speed of $800\, ms/line$.}
	\label{Fig3}
\end{figure}

To establish the stability and robustness of our sensors, we carried out experiments with a modified commercial low temperature ($4.5\, K$) STM/AFM (Omicron). Tips were first cleaned by a series of e-beam heating, field emission and field evaporation in a field ion icroscope (FIM). To avoid destroying the micro-tip or splitting it from the prong, it is important to ebeam for short period of time and relatively low current. We typically ebeam during 15 seconds several times with the tip at $500\,V$ and $1\,mA$. Figure \ref{Fig3}-a shows a typical FIM image of the tip apex of a QPS after the cleaning treatment\cite{Rezeq.2006,Falter.2013}. Additionally, the micro-tip can be further sharpened by FIM nitrogen etching down to a single atom tip\cite{Rezeq.2006,Urban.2012,Pitters.2013}.

Shortly after the cleaning procedure, the sensor is introduced inside the SPM scanner. Figure \ref{Fig3}-b shows resonance  curve at $4.5\, K$  of the QPS used to acquire STM/AFM images presented in figure \ref{Fig2}-c. From different measurements done on different sensors, the quality factor of the FIB fabricated QPS seems to be limited only by the gluing of the fixed prong\cite{Giessibl.2000}. However, most of the sensors exhibit a quality factor of more than 80k, which compares favorably to commercially available QPS where quality factors usually ranges between $10\,k$ and $40\,k$. This along with the high resonance frequency of our FIB fabricated sensors allow for improved imaging sensitivity and scan rates providing consistent and enhanced results. 

Silicon Si(111) was prepared by a series of resistive flash heating up to $1250^{\circ} C$ with a base pressure in the prep chamber of about $3\,10^{-11}\, torr$\cite{Giessibl.2000}. The approach was first done in STM mode. Figure \ref{Fig3}-c and \ref{Fig3}-d show images in empty (+2.0V) and filled states ($-2.0\,V$) respectively. Scanning was very stable in both junction polarities and the tip showed excellent resistance to harsh tip treatments, e.g. controlled crashes (tip forming) and voltage pulses up to $-6\, V$.

Following stable STM imaging, scanning was switched to NC-AFM mode. Figure \ref{Fig3}-e shows a constant frequency shift image ($-12\, Hz$) of a relatively large area ($(30\times 30)\, nm^2$) obtained at $0\,V$. Figure \ref{Fig3}-f is a higher resolution image ($(5\times 5)\, nm^2$). These preliminary results demonstrate the stability and robustness of our sensor to perform SPM experiments.

\section{Conclusion}
To summarize, we described a new technique to fabricate QPS using FIB. Giving the very small mass load to the free prong, the QPS exhibits high Q factor and resonance frequency which is important for increased force gradient detection, lower oscillation amplitude and higher-order eigenmodes operations. Moreover, the stiffness of the sensor is well defined which allows accurate quantitative interpretation of high resolution AFM images, force spectroscopy and energy dissipation. Future work will focus on acquisition of quantitative AFM measurements as well as optimization of the fabrication technique and addressing the cross-talk problem during simultaneous current and force measurements\cite{Majzik.2012}.

\section*{Acknowledgments} 
We thank Martin Cloutier for technical assistance. 
We gratefully acknowledge financial support from NRC, NSERC and AITF.
\bibliography{Qpp_Bib}
\end{document}